\documentclass{emulateapj}

\usepackage{epsfig}
\usepackage{amsmath}
\usepackage{rotating}
\usepackage{natbib}
\usepackage{enumerate}
\usepackage{multirow}
\usepackage{array}
\usepackage{appendix}

\def\plottwoh#1#2{\centering \leavevmode
\includegraphics[height=.32\columnwidth]{#1} \hfil
\includegraphics[height=.32\columnwidth]{#2}}

\newcommand{\cN}[1]{\mathcal{N}}

\def\gsim{\;\rlap{\lower 2.5pt
 \hbox{$\sim$}}\raise 1.5pt\hbox{$>$}\;}
\def\lsim{\;\rlap{\lower 2.5pt
   \hbox{$\sim$}}\raise 1.5pt\hbox{$<$}\;}

\begin{document}


\title{The Deuterium-Burning Mass Limit for Brown Dwarfs and Giant
  Planets}

\author{
David S. Spiegel\altaffilmark{1},
Adam Burrows\altaffilmark{1},
John A. Milsom\altaffilmark{2}
}

\affil{$^1$Department of Astrophysical Sciences, Peyton Hall,
  Princeton University, Princeton, NJ 08544 \\
$^2$Department of Physics, The University of Arizona, Tucson, AZ
  85721}

\vspace{0.5\baselineskip}

\email{
dsp@astro.princeton.edu,
burrows@astro.princeton.edu,
milsom@physics.arizona.edu
}

\begin{abstract}
There is no universally acknowledged criterion to distinguish brown
dwarfs from planets.  Numerous studies have used or suggested a
definition based on an object's mass, taking the $\sim$13-Jupiter mass
($M_J$) limit for the ignition of deuterium.  Here, we investigate
various deuterium-burning masses for a range of models.  We find that,
while $13 M_J$ is generally a reasonable rule of thumb, the deuterium
fusion mass depends on the helium abundance, the initial deuterium
abundance, the metallicity of the model, and on what fraction of an
object's initial deuterium abundance must combust in order for the
object to qualify as having burned deuterium.  Even though, for most
proto-brown dwarf conditions, 50\% of the initial deuterium will burn
if the object's mass is $\sim$$(13.0\pm 0.8)M_J$, the full range of
possibilities is significantly broader.  For models ranging from
zero-metallicity to more than three times solar metallicity, the
deuterium burning mass ranges from $\sim$11.0~$M_J$ (for 3-times solar
metallicity, 10\% of initial deuterium burned) to $\sim$16.3~$M_J$
(for zero metallicity, 90\% of initial deuterium burned).
\end{abstract}

\keywords{radiative transfer -- stars: low-mass, brown dwarfs --
  stars: evolution}

\section{Introduction}
\label{sec:intro}
The year 1995 heralded both the first unambiguous detection of a brown
dwarf \citep{oppenheimer_et_al1995} and the first unambiguous
detections of planets beyond our solar system \citep{mayor+queloz1995,
  marcy+butler1996}.  Many of the first substellar objects detected
were either clearly brown dwarfs (very massive, not in a close orbit
around a main-sequence star) or clearly planets (lower mass, orbiting
close to their stars).  However, as the number of discoveries of
substeller objects grew to the dozens and then hundreds, there
increasingly appeared to be overlap in the apparent mass distributions
of brown dwarfs and planets.  This highlighted the need to clarify the
taxonomy.  Moreover, the various definitions have been strained by the
recent discoveries of objects such as CoRoT-3b, a
$\sim$22-Jupiter-mass ($M_J$) object in a close (0.057-AU) orbit
around its star \citep{deleuil_et_al2008} and the directly imaged
objects of masses $\sim$5-15~$M_J$ at tens of AU from HR~8799 and
Fomalhaut \citep{Marois_et_al_2008, Kalas_et_al_2008}.

A commonly used way to classify objects that are $\sim$10-15 times the
mass of Jupiter is by whether they fuse deuterium (D) in their deep
interiors.  This criterion was adopted in 2002 by the Working Group on
Extrasolar Planets of the International Astronomical Union
\citep{boss_et_al2007}:
\begin{quote}
{\it 1) Objects with true masses below the limiting mass for
  thermonuclear fusion of deuterium (currently calculated to be 13
  Jupiter masses for objects of solar metallicity) that orbit stars or
  stellar remnants are ``planets'' (no matter how they formed). The
  minimum mass/size required for an extrasolar object to be considered
  a planet should be the same as that used in our Solar System.\\
2) Substellar objects with true masses above the limiting mass for
thermonuclear fusion of deuterium are ``brown dwarfs,'' no matter how
they formed nor where they are located.\\
3) Free-floating objects in young star clusters with masses below the
limiting mass for thermonuclear fusion of deuterium are not ``planets,''
but are ``sub-brown dwarfs'' (or whatever name is most appropriate).}
\end{quote}
Although we (and others) do not necessarily endorse the
``deuterium-burning edge'' as the most useful delineation between
planets and brown dwarfs\footnote{Indeed, note that the IAU did not
  adopt this criterion.  See, e.g.,\\
  http://astro.berkeley.edu/~basri/defineplanet/Mercury.htm.} (see
\citealt{burrows_et_al2001}, \citealt{chabrier_et_al2005}, and
\citealt{bakos_et_al2010}), it is a commonly used criterion and
warrants further exploration.

Models of brown dwarfs and giant planets have been calculated since
before the first such objects were discovered \citep{kumar1963,
  zapolsky+salpeter1969, grossman+graboske1973, burrows_et_al1993,
  saumon_et_al1994, burrows_et_al1995}.  Such computations depended
heavily on the equation of state (EOS), and were therefore aided by
the publication of a new EOS for H$_2$-He mixtures
\citep{Saumon_et_al_1995} (the so-called ``SCvH'' EOS).  In addition,
models of the emergent radiation from, and of the temporal evolution
of, brown dwarfs were improved by the application of nongray radiative
transfer theory \citep{burrows_et_al1997, burrows_et_al1998b,
  burrows+sharp1999, burrows1999, burrows_et_al2001,
  burrows_et_al2003b, baraffe_et_al2002, baraffe_et_al2003,
  allard_et_al2003, chabrier_et_al2000, chabrier_et_al2000b,
  chabrier+baraffe2000, sharp+burrows2007, barman2008b} and
sophisticated chemical models \citep{fegley+lodders1994,
  lodders+fegley2002, burrows+sharp1999, sharp+burrows2007,
  allard+hauschildt1995}.

The theoretical study of brown dwarfs and massive planets is now a
maturing field that is of particular current interest, given the
increasing pace of discovery of objects in this mass range.  Previous
model calculations have suggested that deuterium burning turns on near
a mass of 13~$M_J$ \citep{burrows_et_al1997, burrows_et_al2001} or at
0.012~$M_\sun$ ($\sim$12.5~$M_J$,
\citealt{chabrier_et_al2005}).\footnote{Calculations such as this
  implicitly assume that objects start with large initial entropies
  (which leads, eventually, to high central temperatures), as is
  expected for bodies that form from the collapse of a cloud of
  molecular gas \citep{marley_et_al2007}.  Note that if the initial
  entropy is low, objects that are significantly more massive could,
  in principle, avoid burning any significant amount of deuterium,
  though formation scenarios such as this are probably not possible.}
An early such calculation was performed by
\citet{grossman+graboske1973}, who claimed the existence of a
deuterium main sequence near $0.012 M_\sun$, although
\citet{dantona+mazzitelli1985} found, using a lower deuterium
abundance, that there is no deuterium main sequence.  The models of
\citet{grossman+graboske1973} used a deuterium fraction that was about
10 times the Galactic value, and the atmosphere treatment was less
sophisticated than the current state-of-the-art.  Although previous
calculations, and community prejudices, have converged on a mass limit
of $\sim$13$~M_J$, and this mass range does provide a reasonable
estimate for the mass at which significant deuterium burning occurs,
it is worthwhile to clarify the precise range of
deuterium-burning-mass predictions under a variety of model
assumptions.  \citet{saumon+marley2008} have contributed to this
endeavor.  Specifically, defining the minimum deuterium-burning mass
as objects that burn 90\% of their initial deuterium in 10~Gyr, they
find 13.1 and 12.4~$M_J$ for the cloudless and cloudy cases,
respectively.  This shows that how one computes the radiative boundary
condition (with or without clouds, and what type of clouds), matters
significantly.

In this paper, we systematically explore how deuterium burning depends
on a several physical properties of objects.  Our goal in this paper
is to point out that the division between objects that burn deuterium
and those that do not is not strictly 13~$M_J$.  One important aspect
of this is that deuterium burning does not ``turn on'' suddenly at a
particular mass.  Even with just a single set of model parameters, the
mass at which 10\% of deuterium burns is different from the mass at
which 50\% or 90\% burns.  This ambiguity is only enhanced when one
examines the range of possible or likely values of the variables under
consideration here (such as helium abundance, initial deuterium
abundance, and metallicity).  We consider various (cloud-free) models
and nuclear burning criteria; in \S\ref{sec:models}, we describe the
models under consideration.  We vary the helium fraction as a proxy
for the effect of varied concentrations of elements heavier than
hydrogen (\S\ref{ssec:mYHe}); we vary the initial fraction of
deuterium (\S\ref{ssec:mYd}); and we vary the metallicity in the
context of its influence on atmospheric opacity and cooling
(\S\ref{ssec:mZ}).  In \S\ref{sec:results}, we describe the influence
that each aforementioned knob has on deuterium burning, and in
\S\ref{sec:disc}, we calculate derivatives of the D-burning mass edge
with respect to each quantity ($Y$, $y_{\rm D,i}$, $Z$).  In
\S\ref{sec:conc}, we conclude by discussing the implications of our
results for the distinction between brown dwarfs and giant planets.

\section{Models}
\label{sec:models}
We consider a range of brown dwarf/planet models.  If the objects are
massive enough, their deep interiors have conditions under which
non-negligible burning of deuterium occurs through the reaction $p+d
\rightarrow \gamma + {\rm ^3He}$ (by far the most important reaction
involving deuterium at the temperatures and densities in the cores of
these objects).  The rate of deuterium burning depends on the
concentration of deuterium, the mass density, and the temperature.
\citet{stahler1988} presents a power-law approximation to this rate in
the range $T \sim 10^6$~K: $\epsilon_{\rm D} \propto \rho {y_{\rm D}}
T^{11.8}$, where $\epsilon_{\rm D}$ is the specific deuterium-burning
rate, $\rho$ and $T$ are central values of density and temperature,
and $y_D$ is the deuterium mixing ratio.  Although this expression
does not, of course, describe the actual deuterium-burning rate used
in our code, it may serve as a rough guide.  We vary the helium mass
fraction ($Y$, which influences the central density), we vary the
initial deuterium mixing fraction ($y_{\rm D,i}$), and we vary the
metallicity of the atmosphere ($Z$, which strongly affects the
opacity).  Our models are described in \S\ref{ssec:tech}-\ref{ssec:mZ}
below and summarized in Table~\ref{ta:models}.  Note that each row of
this table actually corresponds to a family of models with a dense
spacing of masses, ranging from 10 to 20 times Jupiter's mass.

\subsection{Modeling Technique}
\label{ssec:tech}
The basic modeling strategy is the one described in
\citet{burrows_et_al1997}.  Specifically, we self-consistently
calculate the evolution of these objects by matching
radiative-convective atmosphere models onto fully convective
interiors.  Objects are assumed to be born high on the Hayashi track,
with a given high interior entropy and a flat entropy profile.  This
is a reasonable assumption because in Nature these objects become
fully convective and lose their initial conditions (i.e., the entropy
decreases rapidly from its initial high value) within $10^{3-5}$ years
after formation.  This is far shorter than the duration or age of
deuterium burning, which can last $10^{7-9}$ years.  Interior
structure is calculated with the SCvH EOS \citep{Saumon_et_al_1995}.
In the atmosphere, convective heat flux is treated with mixing length
theory \citep{mihalas1978}, taking the mixing length to be the
pressure scale height.  \citet{burrows_et_al1989} investigated how
other choices of mixing length affect the thermal evolution of brown
dwarfs and low-mass stars and found that, for brown dwarfs, unlike for
stars on the main sequence, this dependence is very weak.

At temperatures found in the interiors of these objects, the rate of
proton-deuteron fusion is reported in the literature as the standard
extrapolation from the rates at higher temperatures.  We use the rate
reported by \citet{caughlan+fowler1988}, which is the same as that
found in \citet{harris_et_al1983}.  In 1999, the Nuclear Astrophysics
Compilation of REaction rates (NACRE) released a new compilation of
thermonuclear rates \citep{angulo_et_al1999}, which was updated
several years later \citep{descouvemont_et_al2004}.  Still, the recent
release of the Reaclib Database \citep{cyburt_et_al2010} by the Joint
Institute for Nuclear Astrophysics (JINA) uses the
\citet{caughlan+fowler1988} rate.
A good discussion of nuclear fusion rates can be found in
\citet{adelberger_et_al2010}.
The \citeauthor{caughlan+fowler1988} rate and that of
\citet{descouvemont_et_al2004} differ by $\sim$20\% at $T\sim 10^8$~K,
and start to converge at lower temperatures.  The interiors of these
substellar-mass objects reach temperatures of
$\sim$0.5-1$\times$$10^6$~K, where the Gamow peak energies are
$\sim$0.7-1~keV.  At these temperatures and energies, the discrepancy
between the older and the newer estimates of the rate is less than
10\%, although extrapolations in this regime are generically
problematic.  To our knowledge, there is no newer information on how
properly to make this extrapolation to temperatures and Gamow peak
energies of relevance to the question at hand.\footnote{The ambiguity
  in the proper prescription for deuterium fusion, of course,
  translates into some ambiguity in the deuterium-burning mass beyond
  what is suggested by the parameter study described in this paper;
  this effect, however, is small.}  Screening corrections for both
ions and electrons are employed \citep{saumon_et_al1996,
  dewitt_et_al1996, sahrling+chabrier1998}.

The thermal evolution depends on the cooling rate as a function of
object size and entropy.  In order to calculate this, as in
\citet{burrows_et_al1997}, we pre-calculate a large grid of one
dimensional, non-gray, radiative-convective atmosphere models with
different effective temperatures ($T_{\rm eff}$) and surface gravities
($g$).  The effective temperature specifies the net
radiative-convective flux at every vertical level ($\sigma T_{\rm
  eff}^4$, where $\sigma$ is the Stefan-Boltzmann constant).  Each
model in the grid is fully convective at its base.  The grid of
atmosphere calculations, therefore, can be thought of as defining a
function $S[T_{\rm eff}, g]$, where $S$ is the entropy.  What is
important for the thermal evolution is the net cooling rate
(essentially $T_{\rm eff}$).  Therefore, we invert this function to
obtain $T_{\rm eff}[S,g]$.  This function depends on the atmospheric
opacities and is what is meant by the ``surface boundary condition.''
At every step of the thermal evolution, the cooling rate is calculated
by interpolating in the grid.  In this way, the fully convective
interior matches smoothly onto the appropriate atmospheric boundary
condition.

\subsection{Helium Fraction}
\label{ssec:mYHe}
For an ideal gas, density depends on molecular weight, while for
degenerate matter, density is inversely related to the ratio of
electrons to baryons.  The abundance of heavier-than-hydrogen elements
affects the density for both reasons.  We vary the helium content of
our models from $Y=0.22$ to $Y=0.32$, with more helium-rich models
representing objects that are richer in both helium and metals.  The
corresponding model names (in Table~\ref{ta:models} and in the
figures) are He22 -- He32.  Model He25 has helium mass-fraction 0.25,
and is also referred to as model D2 (described in \S\ref{ssec:mYd}).
These models have a radiative boundary condition (i.e. atmospheric
opacities) calculated using the method of \citet{burrows_et_al1997},
with solar-metallicity opacities and initial deuterium number fraction
of $y_{\rm D,i} = 2\times 10^{-5}$.

We note that not all helium abundances are equally likely.
Historically, quite low values of $Y$ have been inferred for Jupiter
and Saturn.  Using two different instruments on {\it Voyager},
\citet{gautier_et_al1981} estimated that the helium mass-fraction of
Jupiter was $0.19 \pm 0.05$ or $0.21 \pm 0.06$, respectively.  More
recently, \citet{vonzahn_et_al1998} used {\it Galileo} data to find
the helium mass-fraction of Jupiter to be $0.234 \pm 0.005$.  In a
reanalysis of {\it Voyager} data, \citet{conrath+gautier2000} found
the abundance in Saturn to be in the range 0.18-0.25.  One must keep
in mind that, because of differentiation, atmospheric values for these
objects might be lower than bulk abundances.  Furthermore, the {\it
  Wilkinson Microwave Anisotropy Probe} (WMAP) indicates that a fairly
stringent lower bound on the cosmic primordial helium mass fraction is
0.248 \citep{spergel_et_al2007}.  Higher values than this can easily
be explained in some objects from Galactic enrichment, but lower
values might be difficult to explain.  Nevertheless, we examine the
effect of very low values of $Y$, down to 0.22, so as to explore the
dependence of the deuterium-burning edge on $Y$ values that have been
invoked in the last several decades, but the reader should note that
such an extreme value might not obtain in actual substellar objects.

\subsection{Initial Deuterium Fraction}
\label{ssec:mYd}
Big-bang nucleosynthesis calculations, in conjunction with other
observations (CMB, high-$z$ quasars, etc.) suggest that the primordial
D:H ratio is $(2.8\pm 0.2)\times 10^{-5}$ \citep{pettini_et_al2008}.
Our galaxy, however, is somewhat depleted in deuterium, relative to
this value.  Recent observations indicate that the mean mixing ratio
of deuterium in the interstellar medium is roughly $(2.0 \pm 0.1)
\times 10^{-5}$, though the scatter (among different lines of sight)
is a factor of several \citep{prodanovic_et_al2010}.  In our models,
we take initial deuterium fractions of $10^{-5}$, $2\times 10^{-5}$,
and $4\times 10^{-5}$ (mixing ratios).\footnote{Values below this
  range are not rare \citep{prodanovic_et_al2010}.}  The corresponding
model names are D1, D2, and D4.  These models also have a radiative
boundary condition (i.e. atmospheric opacities) calculated using the
method of \citet{burrows_et_al1997}, with solar-metallicity opacities
and a helium mass-fraction of 0.25.

\subsection{Metallicity}
\label{ssec:mZ}
For a non-irradiated object, the metallicity, i.e. the abundance of
elements heavier than helium, is the main determinant of the
atmospheric chemistry and, hence, the opacity.\footnote{Metals affect
  interior opacity too, but at any metallicity brown dwarf interiors
  are fully convective.} Among objects with thick $\rm
H_2$-He-dominated atmospheres, this abundance can range from very low,
to more than 30 times the solar abundance, as is expected for objects
such as Uranus, Neptune, and similar-mass exoplanets
\citep{guillot+gautier2007, spiegel_et_al2010a, lewis_et_al2010,
  nettelmann_et_al2010, madhusudhan+seager2010}.  Still, objects as
massive 10~$M_J$ might generally be expected to have metallicities not
much more than 0.5--1 dex more than solar.

In our models, we take metallicity values ranging from zero up to a
half dex more than solar, and we calculate the opacity corresponding
to equilibrium chemistry \citep{burrows+sharp1999, sharp+burrows2007}.
Here, we take solar metallicity to be defined by
\citet{anders+grevesse1989}.  The typical derived atmospheric
molecular opacities and, hence, the net cooling rate resulting from
these older abundances generally do not differ much from those
calculated using the more recent \citet{asplund_et_al2009} values, and
the latter remain controversial.  Furthermore, we point out that using
the older elemental abundances is no obstacle to our basic goal, which
is to explore the broad consequences of different metallicities on
thermal evolution.  Since other stars have different elemental ratios
than the Sun, the degree of uncertainty in how our mix of elements
matches that of typical stars depends at least as much on variability
in typical stellar values as on the estimate of solar abundances.  In
the zero-metallicity models, atmospheric opacity results entirely from
collisionally-induced absorption due to the collisionally-induced
dipoles of H$_2$ and He \citep{borysow_et_al1989}.  Though the bottom
end of this range (0 and 0.01 times solar) includes lower
metallicities than might not be expected in most brown dwarfs, it is
interesting to consider the full range of possibilities.

The models in which we vary metallicity fall into two sequences,
Z0.01, Z0.1, Z0.3, Z1.0, and T0.3, T1, T3.  Models in the former
sequence are calculated using the opacities of
\citet{allard+hauschildt1995} to derive the atmospheric boundary
conditions (D. Saumon, private communication) that drive the
evolution.  These models have a helium mass-fraction of 0.25 and
initial deuterium fraction of $2\times 10^{-5}$.  The metallicity of
these models (as a fraction of solar) is the number following the Z in
the model name.  Those in the latter sequence have a boundary
condition calculated using {\tt COOLTLUSTLY} \citep{hubeny1988,
  hubeny+lanz1995, hubeny_et_al2003, burrows_et_al2006,
  sharp+burrows2007}, with the same helium and initial deuterium
fractions \citep{heng+burrows2010}.  Among these, the metallicities of
T0.3 and T3 are not exactly 0.3 and 3, but instead $10^{-1/2}$ and
$10^{1/2}$ times solar metallicity.  Finally, model Z0 has zero
metallicity and is grouped by label with the former sequence, but
actually has a radiative boundary condition taken from
\citet{saumon_et_al1994}.

\section{Results}
\label{sec:results}
By considering a range of model parameters ($Y$, $y_{\rm D,i}$, $Z$)
and atmospheric radiative boundary conditions, and by examining
evolutionary trajectories for a range of masses, at ages from 1~Myr to
10~Gyr, we have produced a broad range of model calculations.  Here,
we show how properties such as radius, effective temperature, and the
deuterium burning rate and power vary with time as a function of mass
(\S\ref{ssec:evol}).  Furthermore, we examine how these evolutionary
tracks vary with the other parameters at fixed mass
(\S\ref{ssec:depend}), and we calculate the variation of the
deuterium-burning mass limit with the model parameters listed above
(\S\ref{ssec:edge}).

\subsection{Evolution with Mass}
\label{ssec:evol}
\begin{figure}[t!]
\plottwoh
{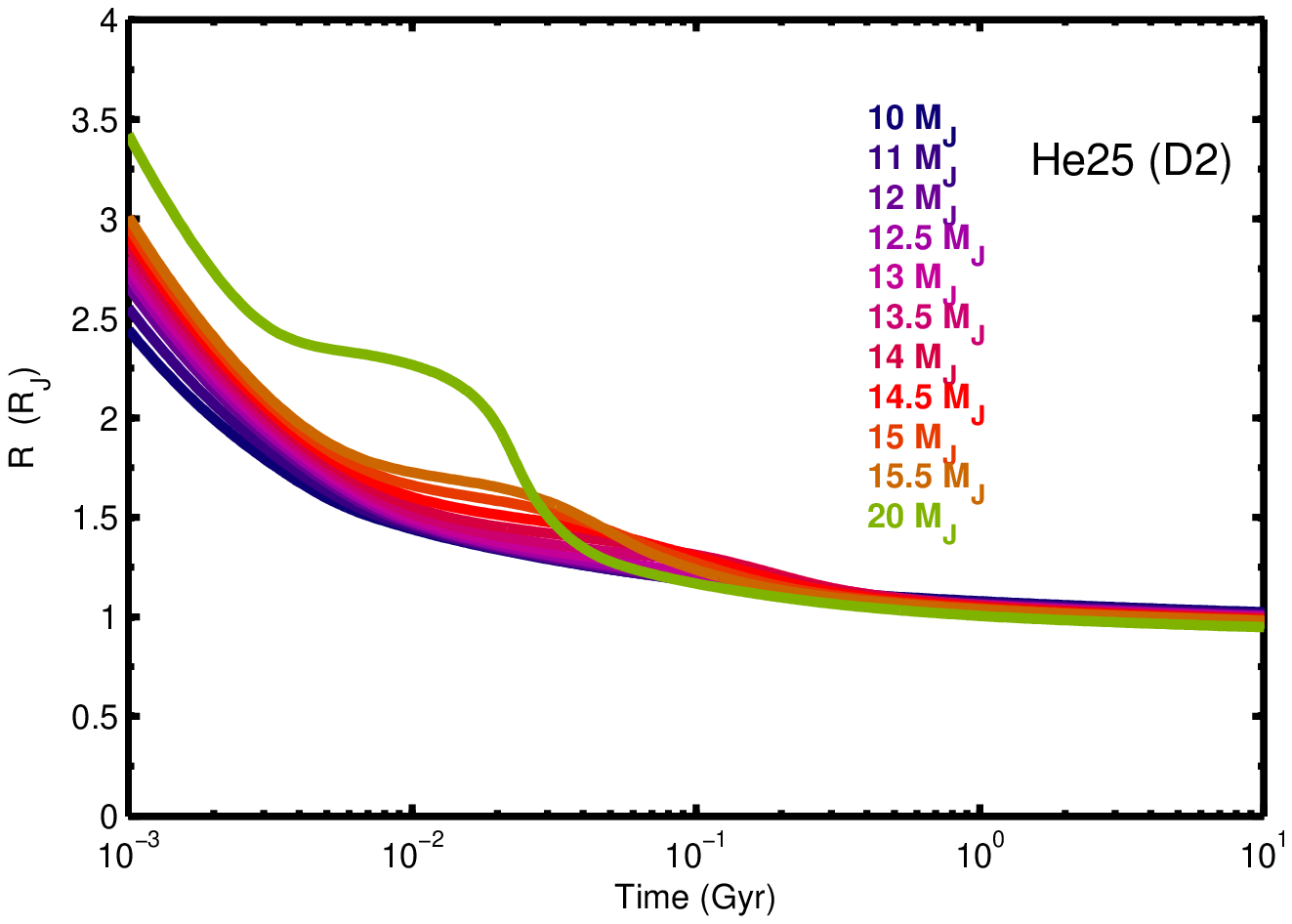}
{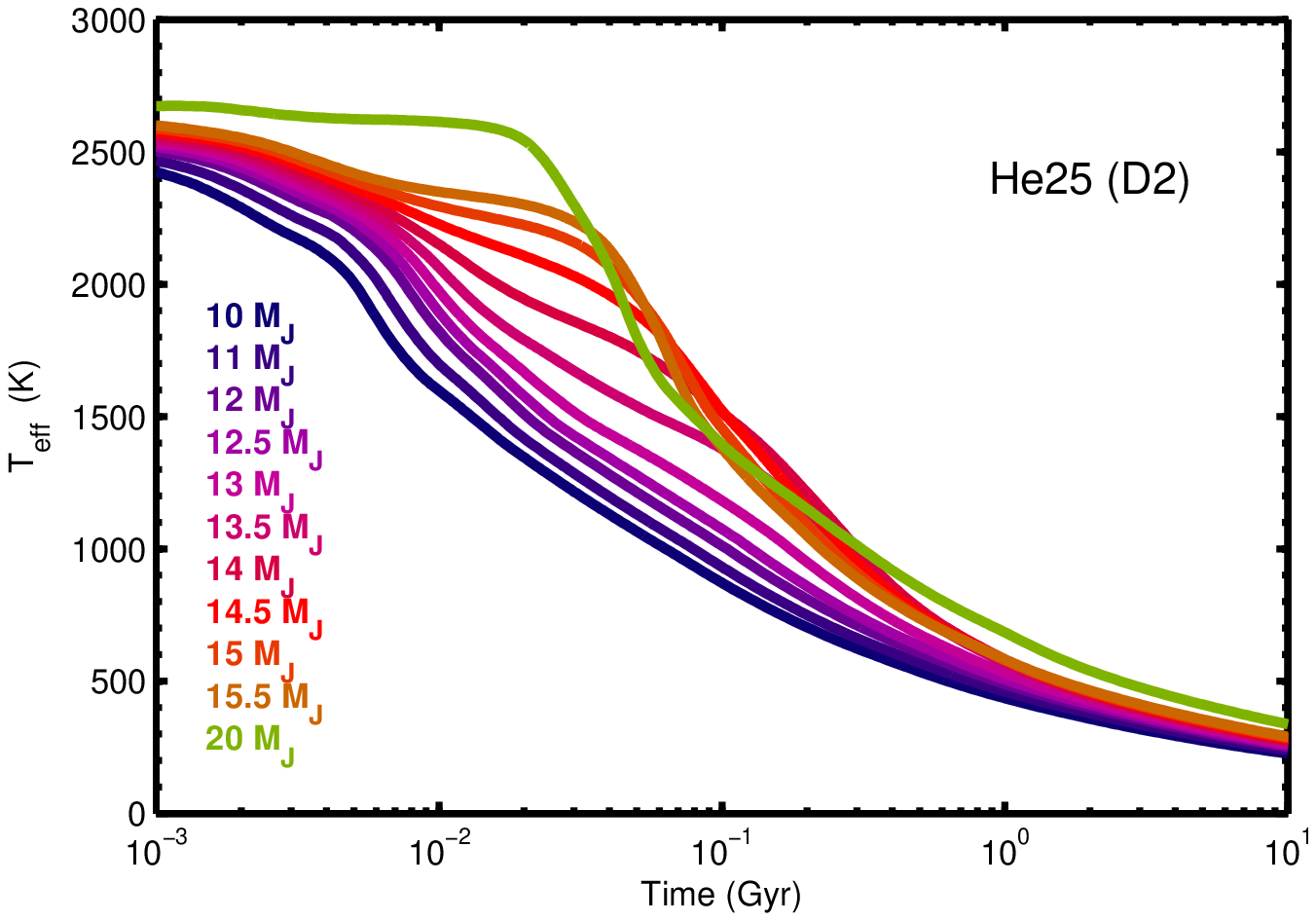}\\
\plottwoh
{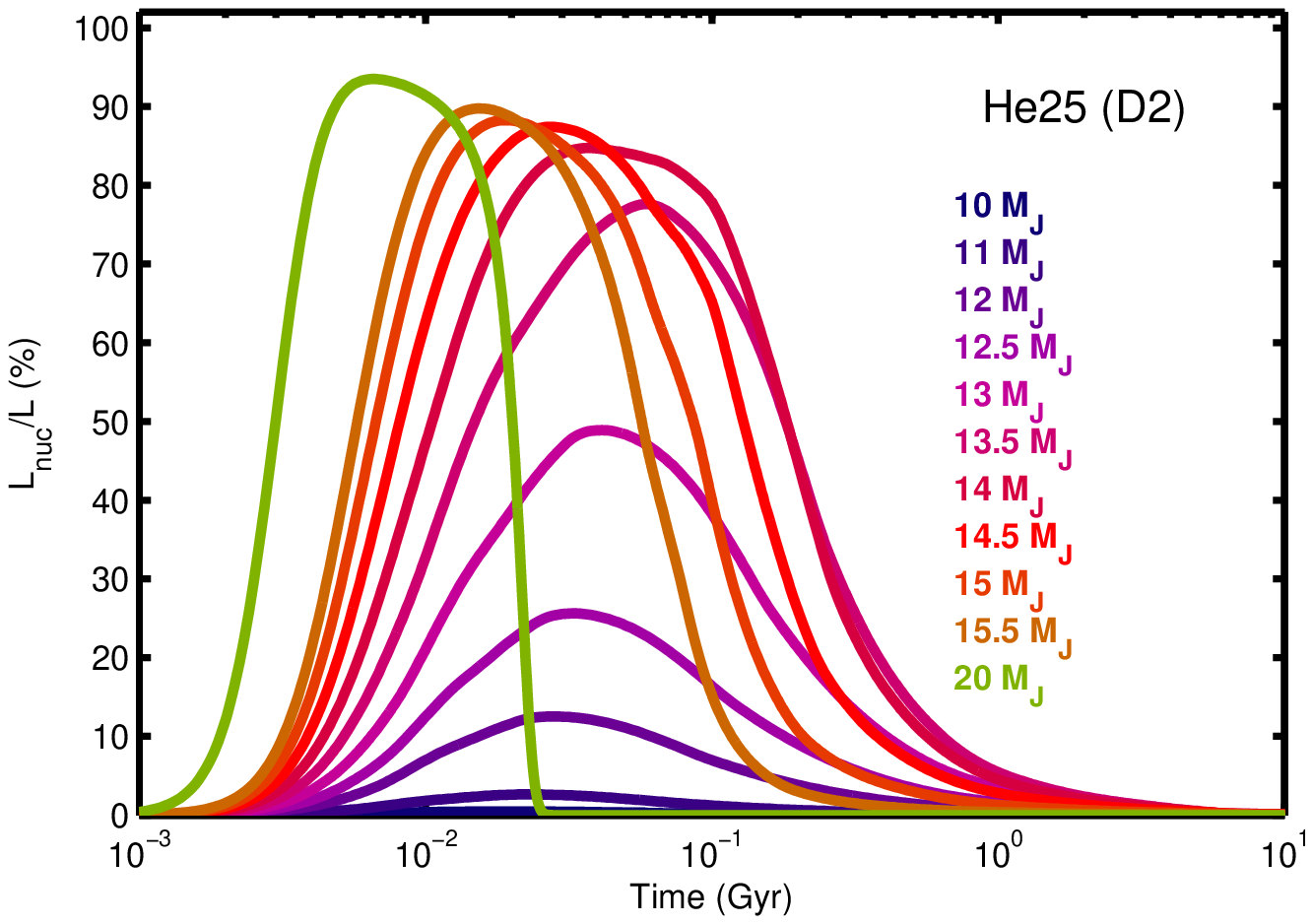}
{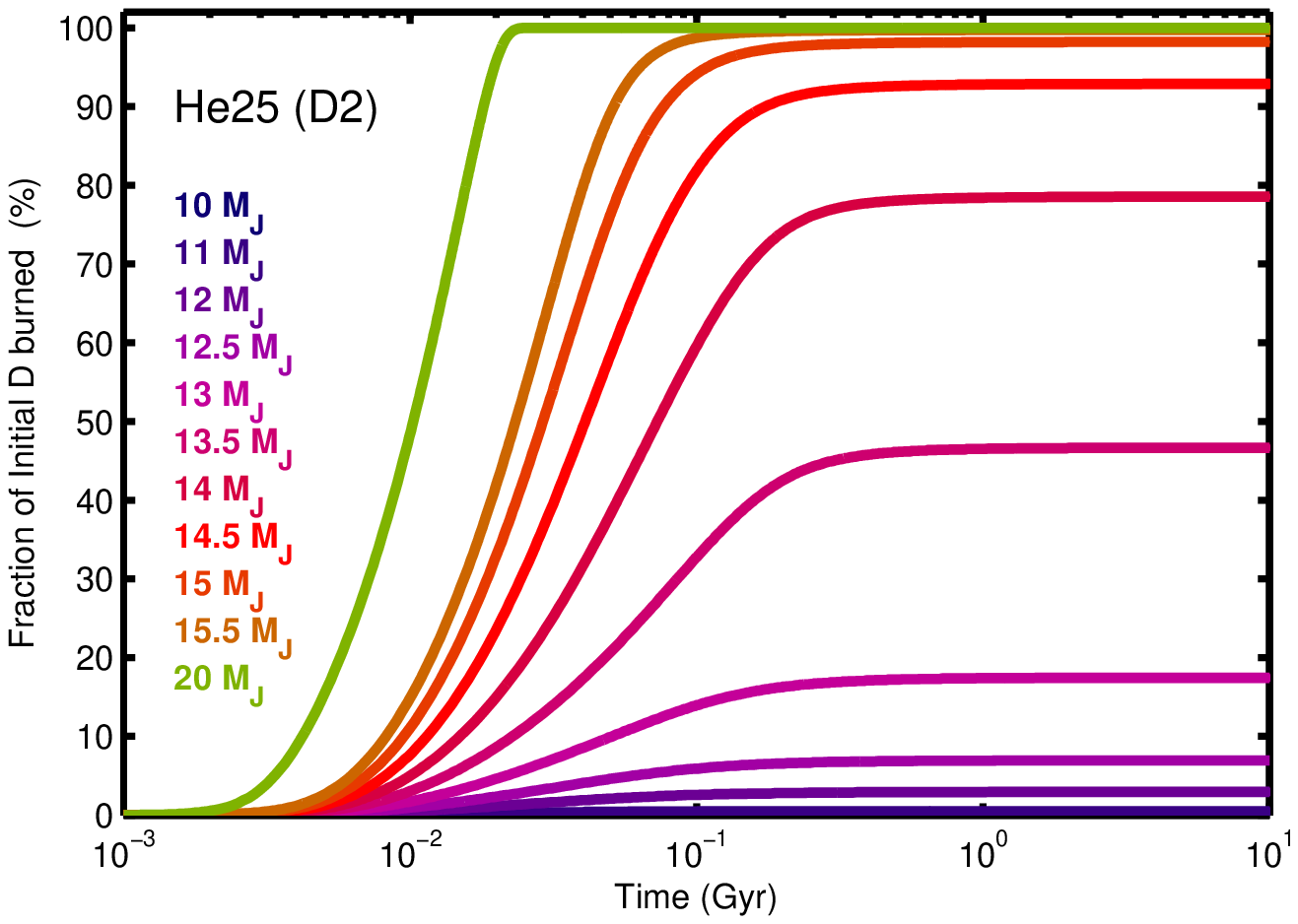}
\caption{Evolutionary trajectories as functions of mass for D2 models.
  {\it Top left:} Radius vs. time.  Higher mass objects cool faster
  and have smaller radii by 10~Gyr, even though they begin with larger
  radii.  The 15-$M_J$ and 20-$M_J$ models cool and shrink
  particularly rapidly after the conclusion of deuterium burning. {\it
    Top Right:} Effective temperature vs. time.  {\it Bottom left:}
  Ratio of nuclear power to total luminosity vs. time.  {\it Bottom
    right:} Fraction of initial deuterium burned vs. time.}
\label{fig:diagnosticsB2}
\end{figure}

The basic character of brown dwarf cooling trajectories is well known
from a number of calculations in the last several decades, but it is
worthwhile to examine the specific behavior of our models.  We begin
with an illustrative set of evolutionary trajectories for a
representative family of models (He25, or D2).
Figure~\ref{fig:diagnosticsB2} shows the evolution of four quantities
that our models track.

The top panels show quantities related to model cooling.  The top left
panel shows the evolution of radius with time, and the top right panel
shows the evolution of effective temperature with time.  It is
instructive to examine these two panels together.  Higher mass objects
begin larger but have higher effective temperatures and therefore cool
more rapidly.  As a result of their faster cooling, the radii of more
massive objects shrink more rapidly, eventually (after roughly a
gigayear) overtaking lower mass objects in radius.  By the end of the
calculation (after 10~Gyr), all objects are very nearly Jupiter's
radius, but more massive objects are slightly smaller.

The lower panels of Fig.~\ref{fig:diagnosticsB2} are related to the
fusion of deuterium.  During the initial Hayashi-track stages of
evolution, the luminosity is dominated by Kelvin-Helmholtz
contraction.  After a few million years, the central part of the
object reaches temperatures and densities at which the deuterium
fusion rate is non-negligible.  The bottom left panel shows the ratio
of the power from nuclear (deuterium) burning to the total luminosity.
For the most massive objects, this ratio reaches fairly large values
(well over 80\%).  However, the main sequence is where this ratio is
essentially 100\% (i.e., where Kelvin-Helmholtz contraction has
ceased), and so it is clear that none of these models goes through a
deuterium main sequence stage.  The bottom right panel shows the
evolution of the fraction of the initial deuterium (in this case,
$y_{\rm D,i}=2\times 10^{-5}$) that has burned.  Higher mass objects
burn a larger fraction of their initial deuterium, and do so faster
than lower mass objects.  Regardless of mass, very little deuterium
burns after a few hundred million years.  Note, in comparing the top
row to the bottom row, that the 15-$M_J$ and 20-$M_J$ models, which
fuse the greatest amount of deuterium (among the models displayed),
cool and shrink very quickly after the deuterium-burning phase.

\subsection{Model-Dependence of D-Burning Evolution}
\label{ssec:depend}
\begin{figure}[t!]
\plottwoh
{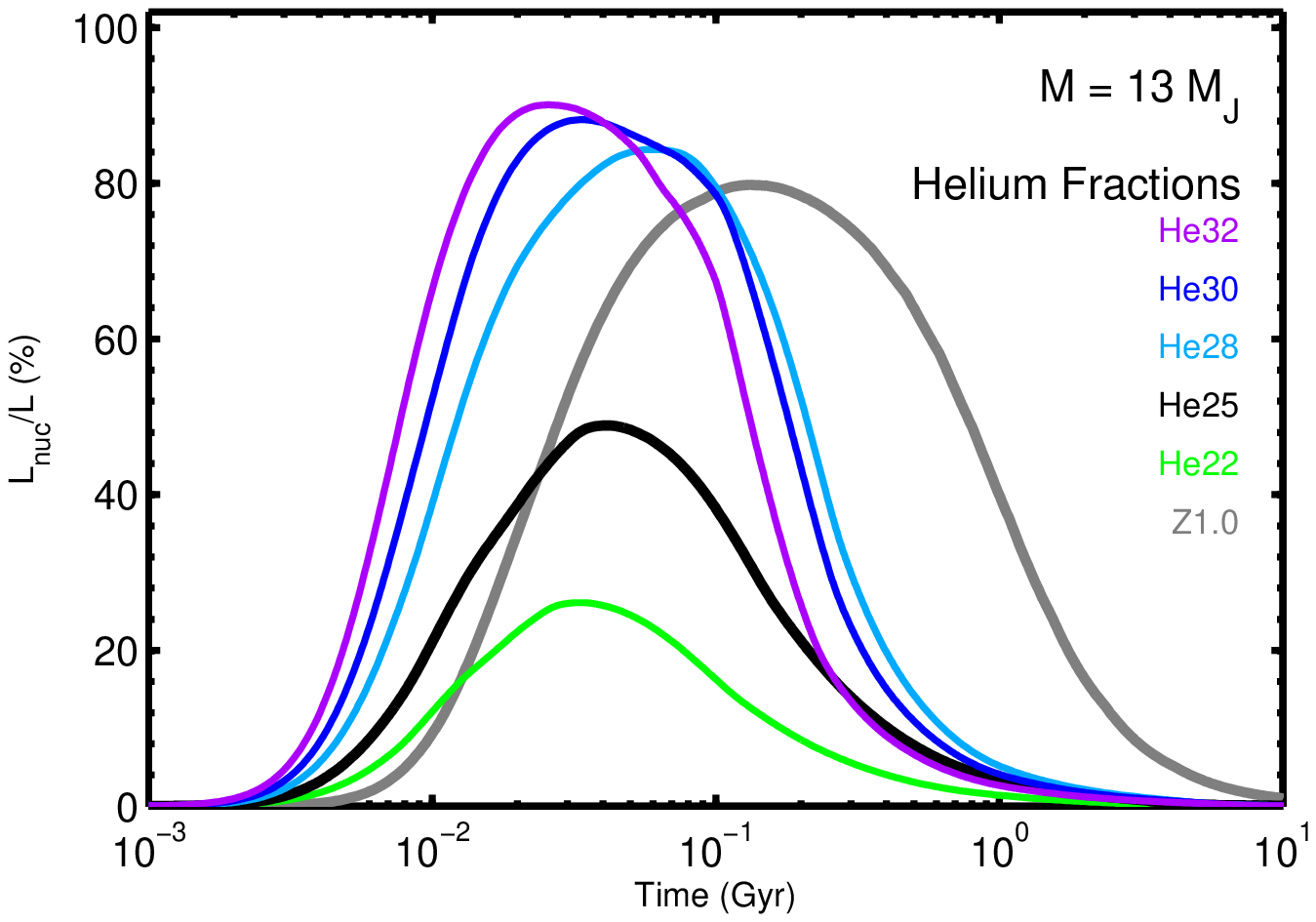}
{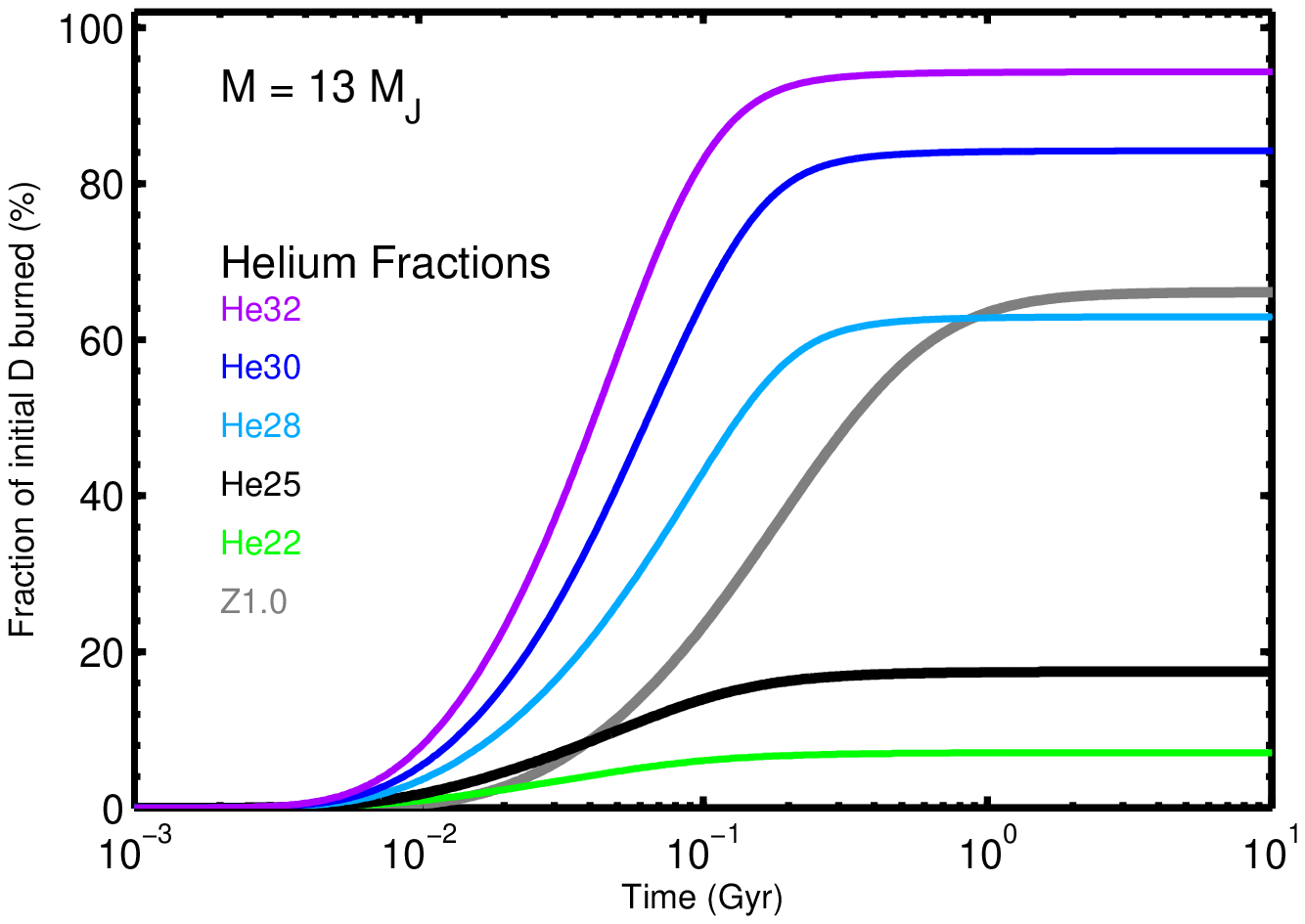}
\plottwoh
{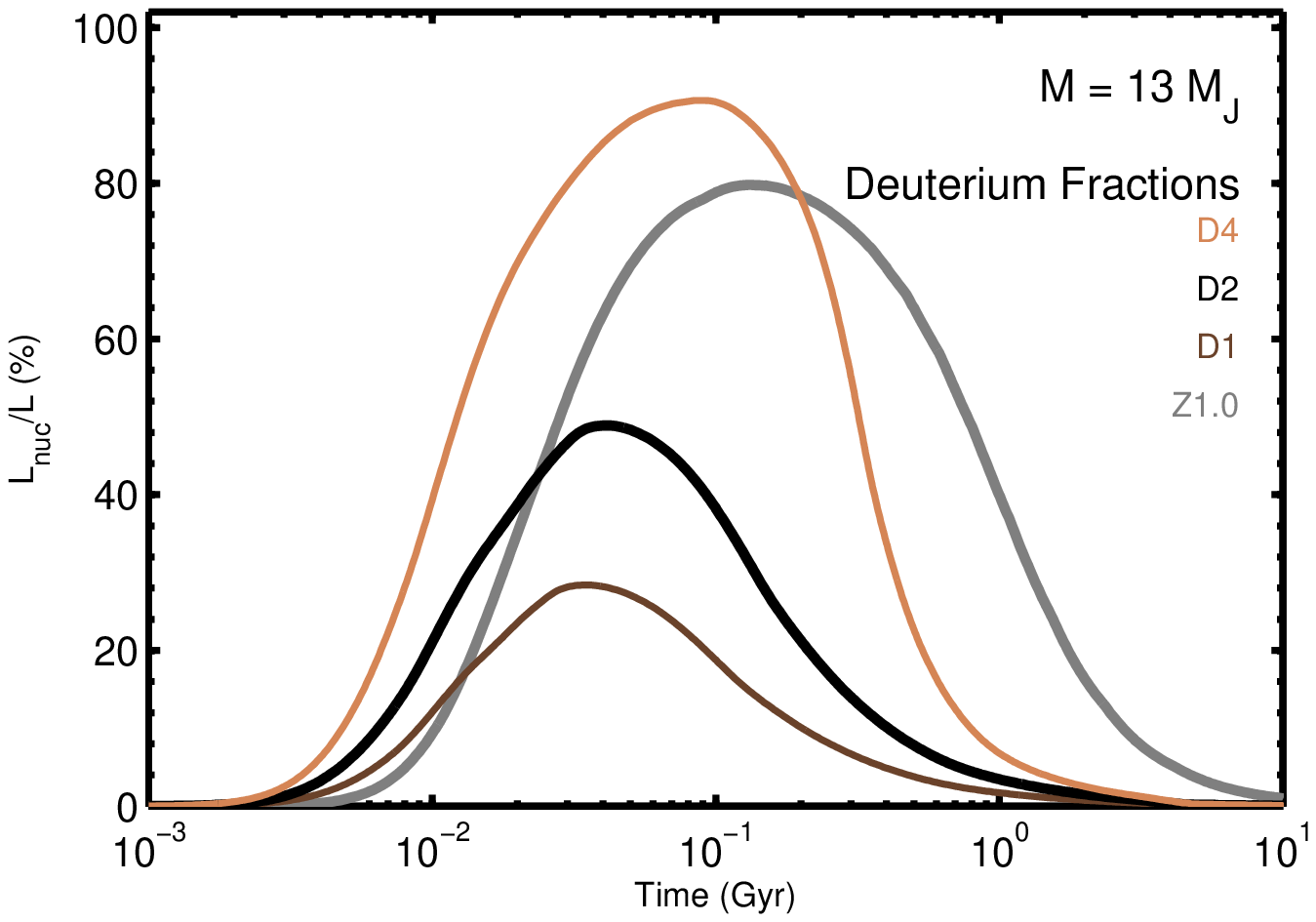}
{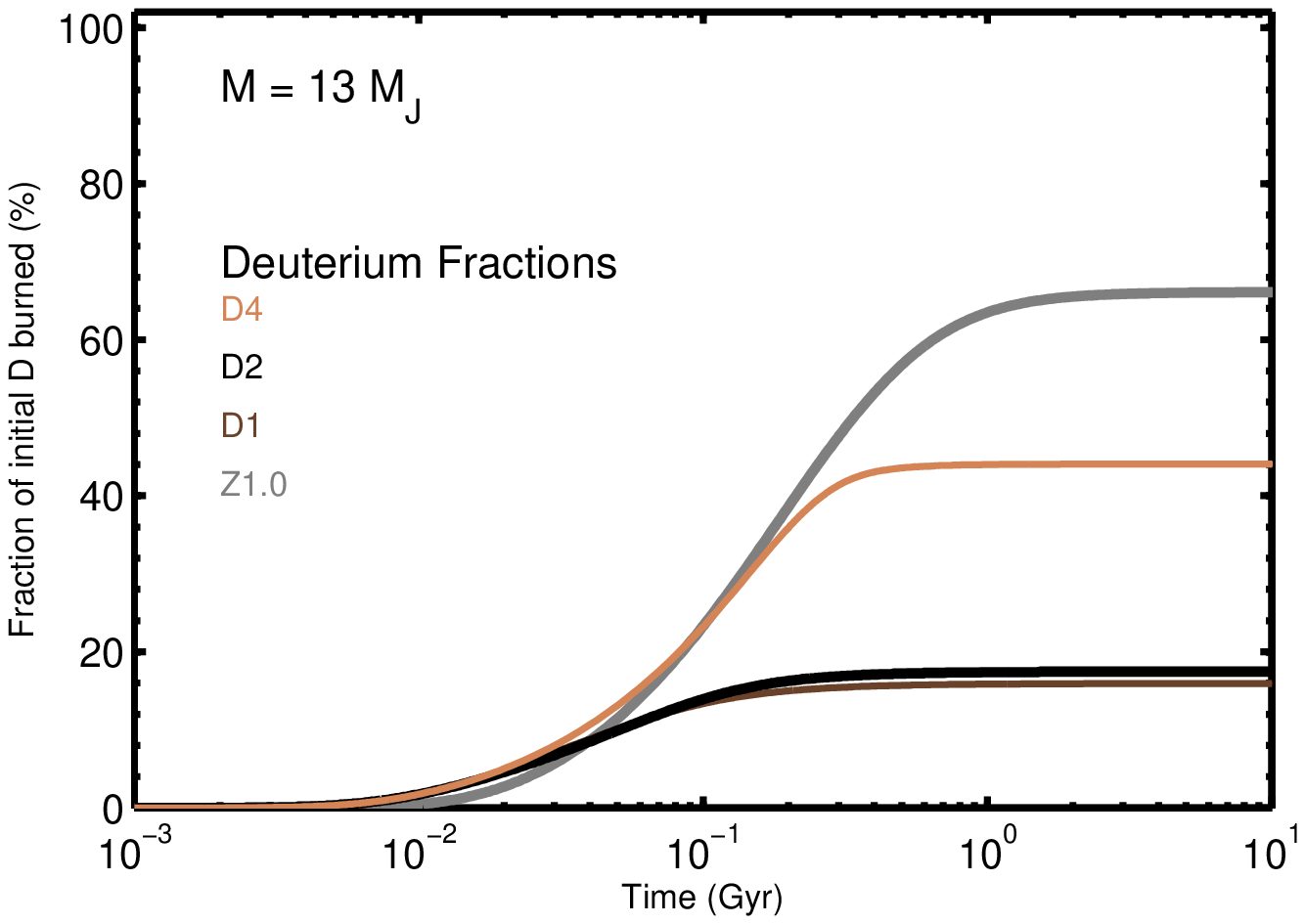}
\plottwoh
{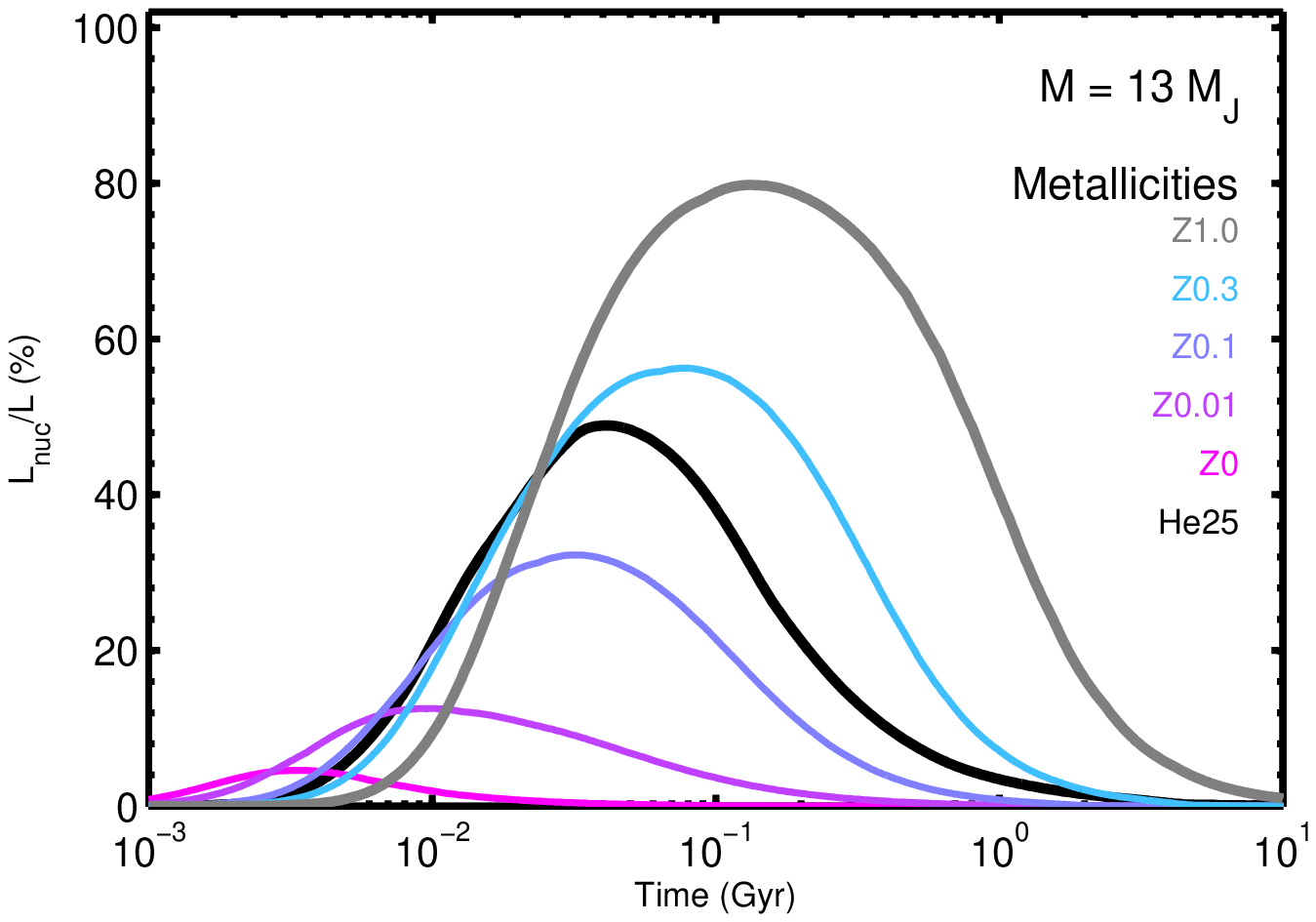}
{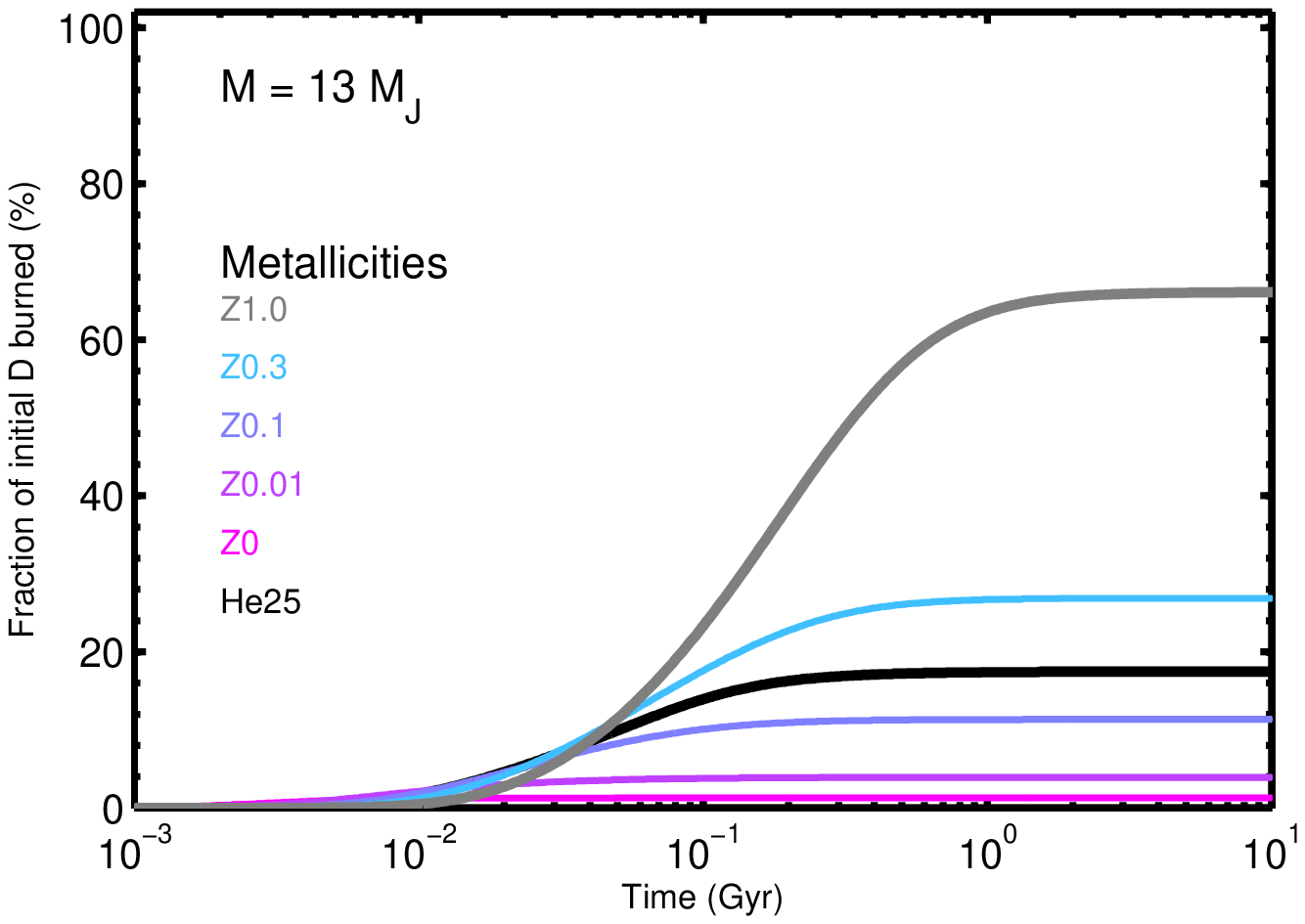}
\caption{Ratio of nuclear power to total luminosity vs. time ({\it
    left}) and fraction of initial deuterium that burns vs. time ({\it
    right}), for 13-$M_J$ models.  Note that very little deuterium
  burning occurs after $\sim$1~Gyr.  Models are grouped by varying the
  helium fraction ({\it top}), the initial deuterium fraction ({\it
    middle}), and the metallicity ({\it bottom}).}
\label{fig:frac_burnedt}
\end{figure}

In Fig.~\ref{fig:frac_burnedt}, we examine the deuterium-burning
evolutionary profiles of 13-$M_J$ objects from a variety of models.
Each row shows analogous plots to the bottom row of
Fig.~\ref{fig:diagnosticsB2}, only instead of different curves
representing different mass objects, they represent different model
properties.  The left column shows the evolution of the ratio of
nuclear power to total luminosity, and the right column shows the
fraction of the initial deuterium that burns.  The three rows show,
from top to bottom, models He22--He32 (varying helium fraction),
models D1--D4 (varying initial deuterium fraction), and models
Z0--Z1.0 (varying metallicity through its influence on atmospheric
opacity).  Each row also shows models He25 (which is also D2) and
Z1.0, for comparison to ``fiducial'' models.

First, it is important to note the difference {\it at fixed model
  parameters} of the radiative boundary condition.  The Z-sequence
models have higher opacity, which leads to slower cooling, than the
other models (which have the \citealt{burrows_et_al1997} boundary
condition).  As a result, models Z1.0 and He25/D2, which are otherwise
identical, have markedly different evolutionary trajectories.  The
former reaches a peak nuclear-to-total power ratio of $\approx$80\%,
while the latter reaches only $\approx$50\%.  Furthermore, the former
eventually burns 66\% of its initial deuterium, while the latter burns
only 17\% of its.  In fact, of the Z-sequence models, model Z0.3 (with
0.316$\times$ solar elemental abundance) more nearly approximates the
He25/D2 evolution than does Z1.0.  With its slower cooling, model Z1.0
has a moderate nuclear-burning to total luminosity ratio until later
times; the ratio shown in the left column stays above 5\% until
$\sim$5~Gyr.

More generally, within each row of Fig.~\ref{fig:frac_burnedt}, the
expected trends hold.  The top row shows that more helium-rich models
burn a greater fraction of their initial deuterium, which is
unsurprising because these are the models with fewer electrons to
support the mass via degeneracy pressure; thus, these models have
denser, hotter cores.  The middle row shows that models with higher
initial deuterium content burn a greater fraction of their initial
deuterium.  This is a somewhat subtle point.  The deuterium-burning
rate is roughly linear in $y_D$, suggesting that, other things being
equal, the fractional depletion of deuterium with time should be
constant irrespective of initial deuterium abundance.  However, other
things are not equal.  In particular, the central temperature of more
deuterium-rich models is maintained at a higher value for a longer
period of time.  As a result, the integrated fractional amount of
deuterium burned is greater for these models.  The bottom row shows
that more metal-rich models burn a greater fraction of their initial
deuterium, which stands to reason because their higher opacities
produce a blanket effect, allowing them to maintain higher core
temperatures for a longer time.

\subsection{Model-Dependence of D-Burning Limit}
\label{ssec:edge}
\begin{figure}[t!]
\plottwoh
{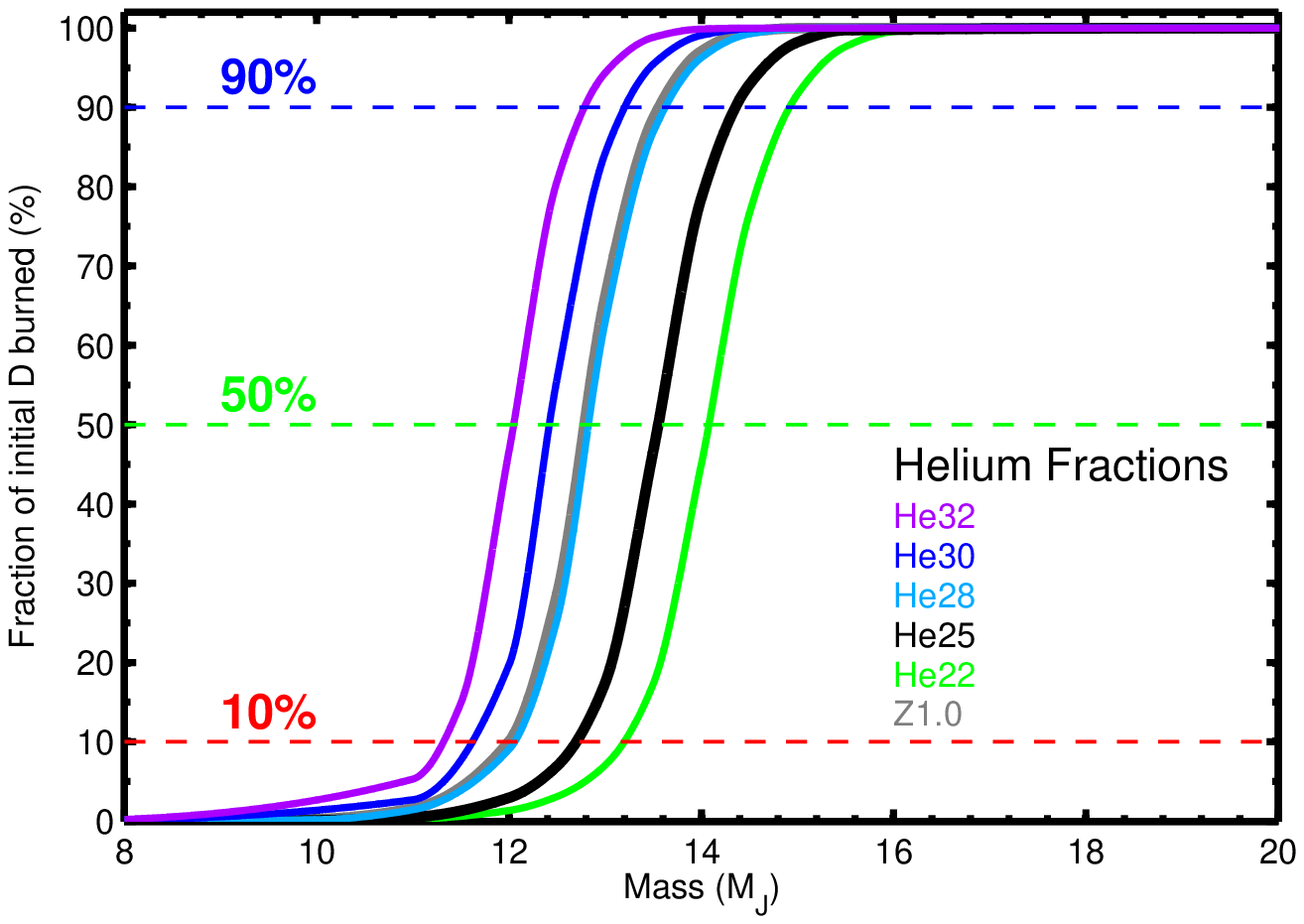}
{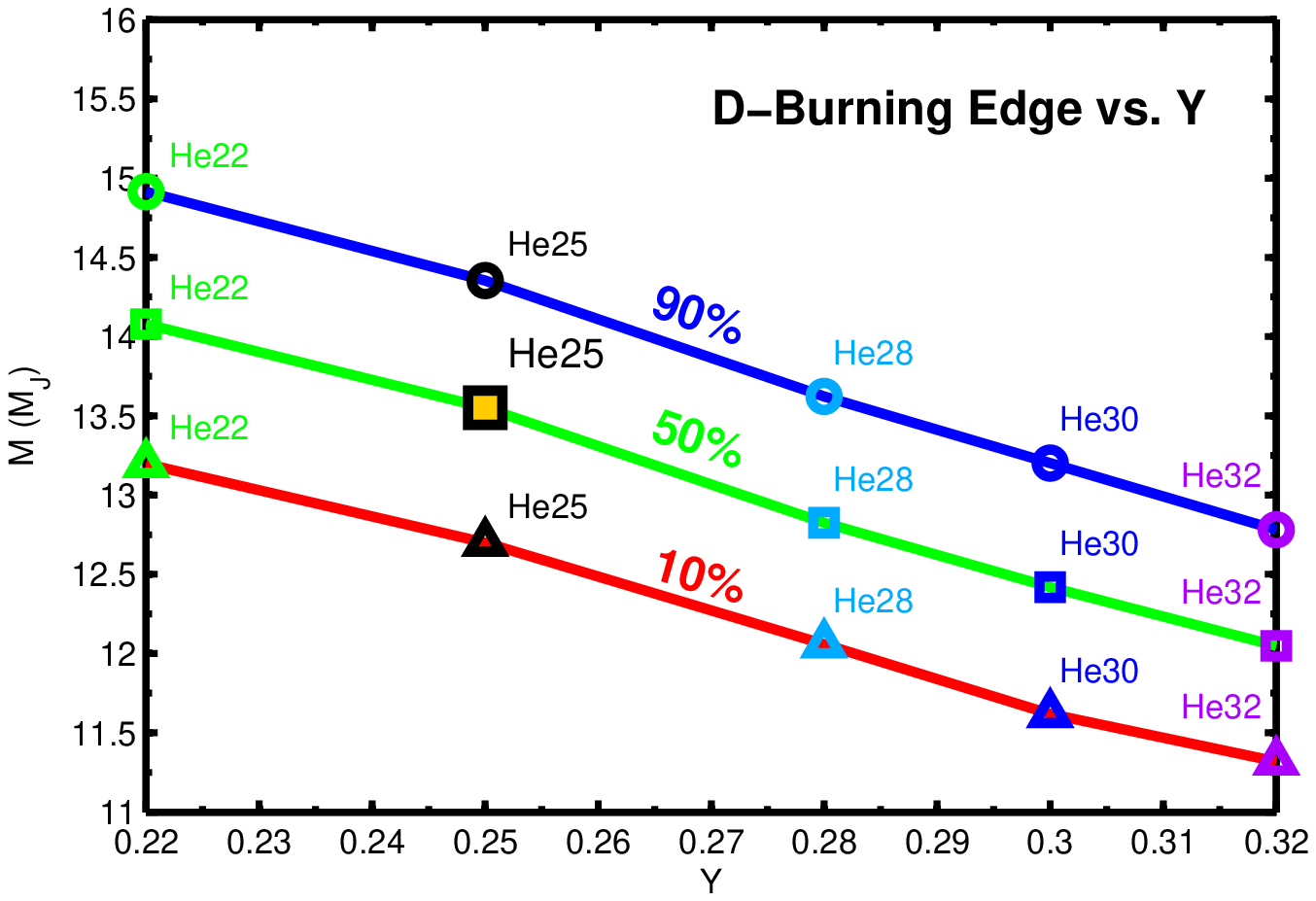}
\plottwoh
{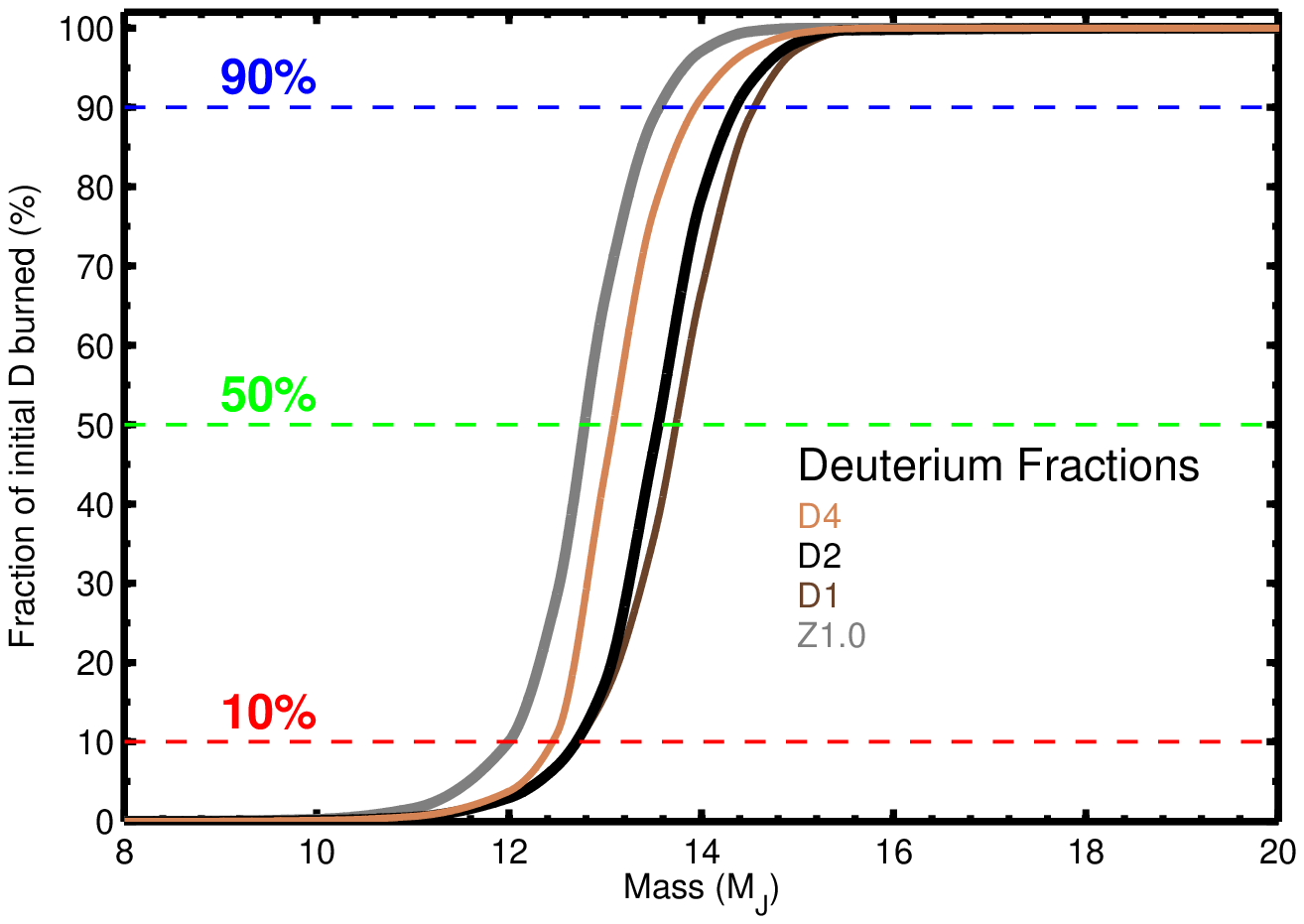}
{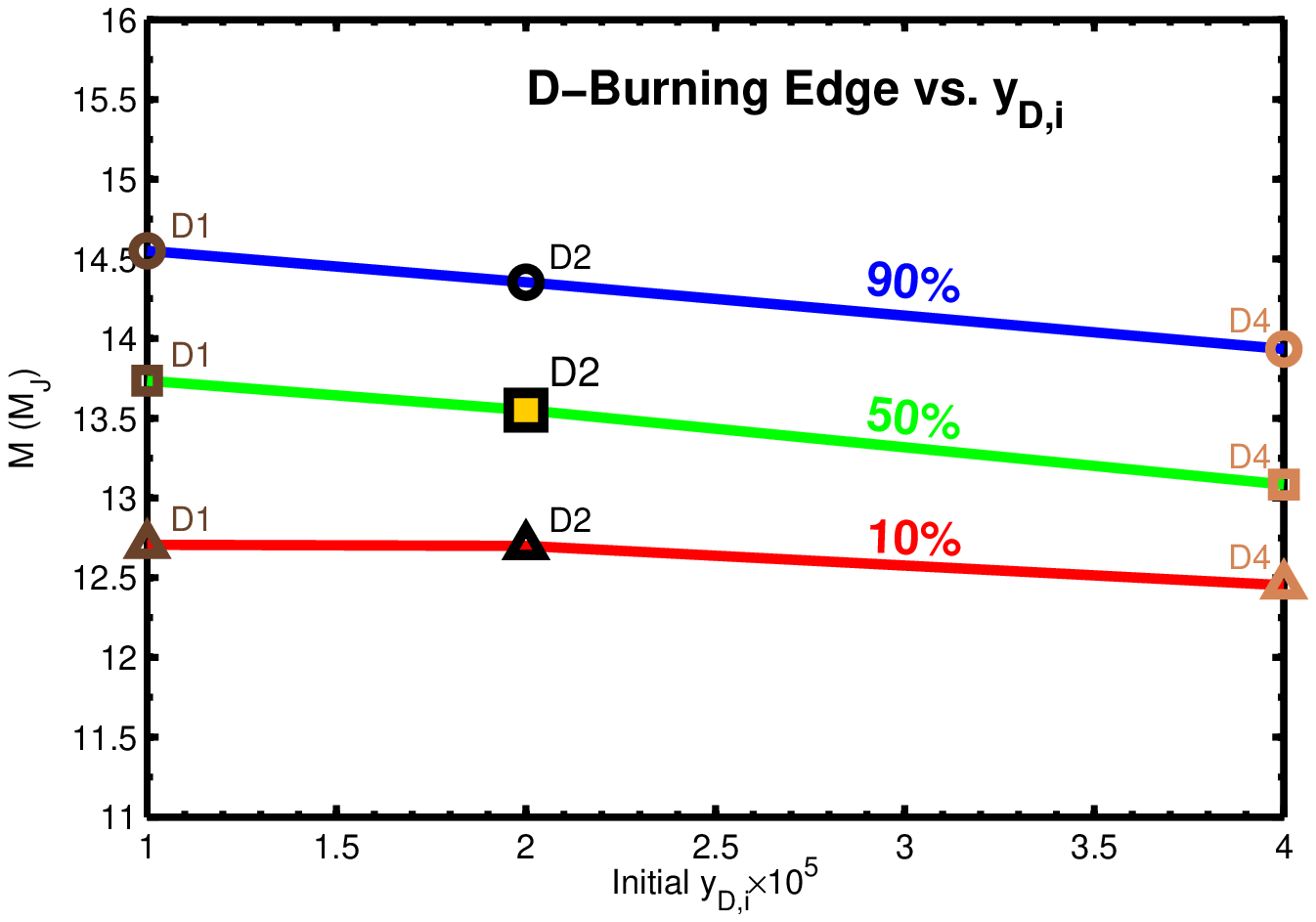}
\plottwoh
{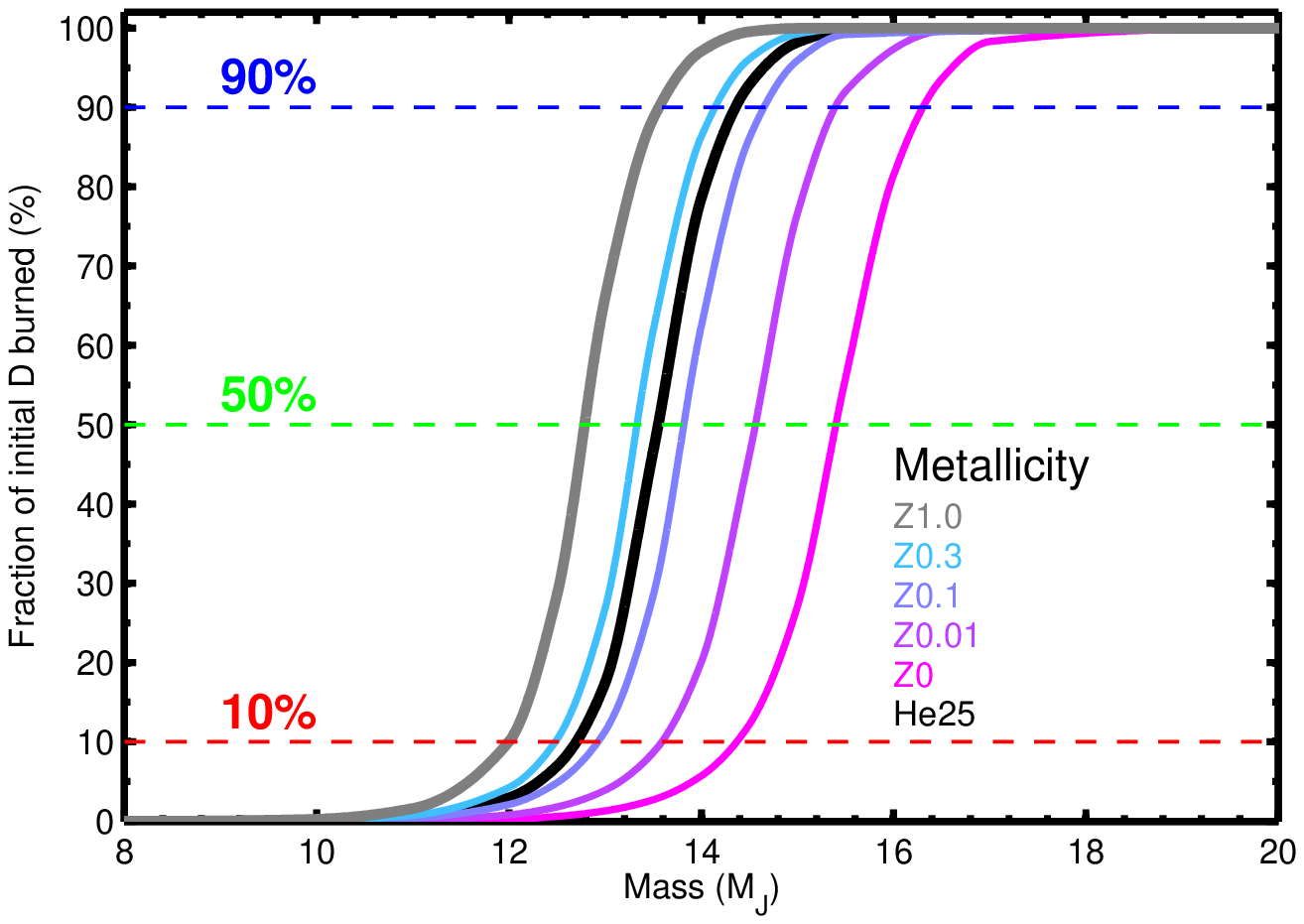}
{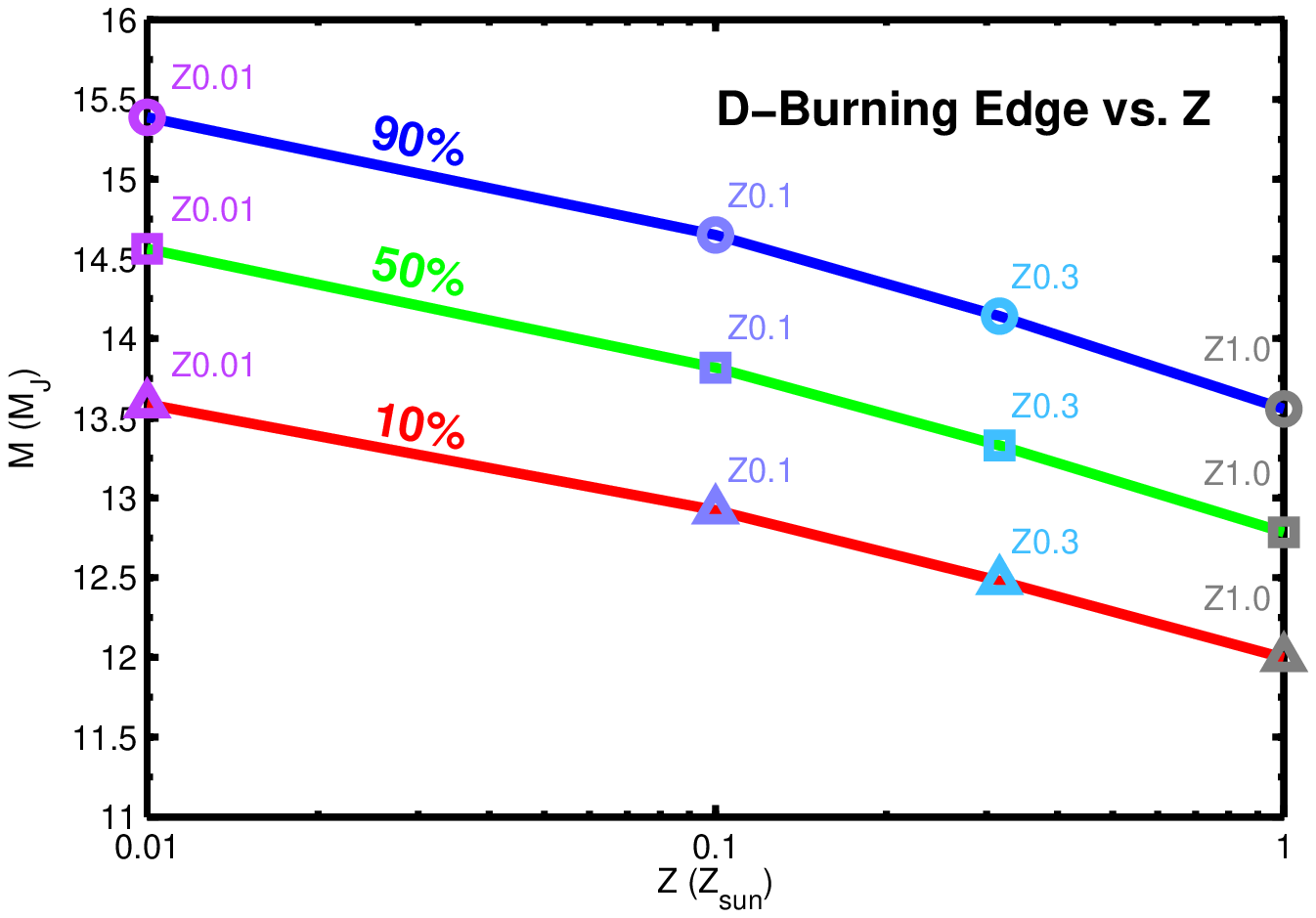}
\caption{Fraction of initial deuterium that burns within 10~Gyr
  vs. Mass ({\it left}); and dependence of deuterium-burning edge on
  $Y$, $y_{\rm D,i}$, and $Z$, for different ``edge-criteria'' ({\it
    right}).\\
  {\bf Left:} For a variety of models, the total fraction of the
  initial deuterium abundance that combusts through nuclear fusion
  within 10~Gyr is shown as a function of the object's mass.  Note,
  per Fig.~\ref{fig:frac_burnedt}, that if any appreciable fraction of
  the initial deuterium ends up burning, this happens within the first
  1~Gyr.  Models He25/D2 and Z1.0 are shown in all three panels.  In
  each panel, horizontal dashed lines are plotted at 10\%, 50\%, and
  90\%.\\ {\it Top:} For solar metallicity and initial deuterium
  number fraction of $2\times 10^{-5}$, models from
  \citet{burrows_et_al1997} are shown for different helium
  mass-fractions.  Greater helium fraction leads to deuterium burning
  at a lower mass.\\
  {\it Middle:} For solar metallicity and helium abundance (by mass)
  of 0.25, models from \citet{burrows_et_al1997} are shown for
  different initial deuterium abundances.  Greater initial deuterium
  abundance leads to deuterium burning at lower mass.\\
  {\it Bottom:} For inital deuterium abundance of $2\times 10^{-5}$
  and helium abundance of 0.25, models using an
  \citet{allard+hauschildt1995} boundary condition (Z0.01, Z0.1, Z0.3,
  Z1.0) are shown.  A model with zero metallicity (Z0,
  \citealt{saumon_et_al1994}) is also shown.  Greater metallicity
  leads to deuterium burning at lower mass.\\
  {\bf Right:} In each panel, the red, green, and blue curves
  correspond to edge-criteria of 10\%, 50\%, and 90\% of the initial
  deuterium burning within 10~Gyr.  These correspond to the masses at
  which the fraction-burned curves in the left panel of this figure
  cross the three horizontal dashed lines.  Individual models are
  coded by the same colors as in the left panel.  The mass of the
  deuterium burning edge is shown as a function of helium fraction
  ({\it top}), initial deuterium fraction ({\it middle}), and
  metallicity ({\it bottom}).  The ``fiducial model'' He25/D2 is
  represented (in the top two panels) with a large black square,
  filled with yellow.}
\label{fig:edges}
\end{figure}

We now quantify the deuterium-burning mass limit, and its dependence
on the various model parameters we have varied.
Figure~\ref{fig:edges} displays the same basic information in two
different ways.  In the left column, the fraction of the initial
deuterium that combusts within 10~Gyr is plotted versus object mass.
Horizontal dashed lines are plotted at 10\%, 50\%, and 90\%,
illustrating various (arbitrary) deuterium-burning cut-offs.  The
right column plots the deuterium-burning mass edge for each of these
three criteria (i.e., the mass at the intersection of each curve in
the left column with the corresponding dashed line) as a function of
the three tunable model parameters: $Y$, $y_{\rm D,i}$, and $Z$.  From
top to bottom, the rows show the same respective model sets as in
Fig.~\ref{fig:frac_burnedt}.  As in Fig.~\ref{fig:frac_burnedt},
models He25/D2 and Z1.0 are shown in all three panels of the left
column.  The last three columns of Table~\ref{ta:models} contain the
same information as the right column of Fig.~\ref{fig:edges}, and also
include data for the {\tt COOLTLUSTY} sequence of models (T0.3--T3).
Note that the solar-metallicity {\tt COOLTLUSTY} model (T1) and the
fiducial \citet{burrows_et_al1997} model (He25/D2) are quite similar
to one another.

There are four main lessons from the data in Fig.~\ref{fig:edges} and
Table~\ref{ta:models}:
\begin{enumerate}
\item Within each sequence of models, the expected trends hold.  That
  is, greater helium abundance, greater deuterium abundance, and
  higher metallicity allow a given amount of deuterium to fuse at a
  lower object mass.
\item Depending on what is meant by ``deuterium burning'' (i.e., how
  much of the initial deuterium must burn to qualify), the
  deuterium-burning mass limit can vary quite a bit.  As the criterion
  goes from 10\% to 50\% and from 50\% to 90\%, the required mass
  increases by $\sim$0.8-0.9~$M_J$.
\item 0.012~$M_\sun$ and 13~$M_J$ is not far from the mass limit for
  most models.
\item However, the full range of masses displayed in this table goes
  from 11.3~$M_J$ (He32, 10\%) to 16.3~$M_J$ (Z0, 90\%).
\end{enumerate}
With the final point in mind, it is clear that claiming that an object
does or does not ``burn deuterium'' implies a set of physical
assumptions about the model and a criterion for deuterium burning.
Any criterion is fine, so long as people know what it is; but implicit
assumptions should be made explicit, and might not actually apply to
some astrophysical objects.

\section{Discussion}
\label{sec:disc}
A useful way to quantify the influence of varying the model parameters
on the deuterium mass limit is with the derivative of the mass-limit
with respect to each model parameter.  Although the curves in the
right column of Fig.~\ref{fig:edges} do not have constant slope, they
are not very far from linear over the ranges shown.  In
Table~\ref{ta:derivs}, we present the average derivative with respect
to each parameter.  For $Y$ and $y_{\rm D,i}$, we take linear
derivatives; for $y_{\rm D,i}$ and $Z$, we take logarithmic
derivatives.  These derivatives are not terribly precise measures of
the variation, but allow for a quick, crude estimate of the deuterium
edge-mass at different parameter values.

Note that, in the calculations discussed so far, the effect of higher
metallicity has been reflected only in the atmospheric opacity.  In a
real object, however, higher metallicity throughout would also lead to
an EOS change for the bulk, because high-$Z$ elements have lower
electron-to-baryon ratios.  As a result, a higher metallicity object
would have a higher central density than occurs in our models, and
would therefore have a lower deuterium mass limit than we have shown,
particularly for model T3.  Though there is no published robust
equation of state that properly includes heavy elements beyond helium,
we can approximately correct for this by using a higher helium
fraction (this was also done by \citealt{guillot2008}).  We,
therefore, have post-processed our models in an attempt to incorporate
this effect.  Using values from \citet{asplund_et_al2009}, we take
$Y=0.2703$ and $Z=0.0142$ to be the helium and heavy-element
mass-fractions corresponding to protosolar abundance.  Motivated by
this, we add 0.0142 in $Y$ for every unit in metallicity (i.e., $Y =
0.2703+0.0142$ for solar, $Y = 0.2703+2\times 0.0142$ for $2\times$
solar, etc.).\footnote{It is not clear that the equivalent $\Delta Y$
  for solar metallicity should be exactly $Z$, but this seems a
  reasonable first approximation.}  We perform a cubic interpolation
in $Y$ and use the $Z$-derivative for model sequence T0.3--T3 to
calculate the values in Table~\ref{ta:redge} corresponding to 0.5, 1,
2, and 3 times solar metallicity.  Values of the edge mass in this
table then range from $\sim$11.0~$M_J$ ($3\times$~solar, 10\% of
initial deuterium burned) to $\sim$13.9~$M_J$ ($0.5\times$~solar, 90\%
of initial deuterium burned).

Although the deuterium-burning mass limits in Tables~\ref{ta:models}
and \ref{ta:redge} vary from 11.0 to 16.3 times Jupiter's mass, a
narrower range of values is found if we restrict our attention to a
particular burning fraction criterion (50\%, say) with realistic $Y$,
$y_{\rm D,i}$, and $Z$ values.  For helium mass-fractions between 0.25
and 0.30, initial deuterium fractions between $10^{-5}$ and $2\times
10^{-5}$, and metallicities between 0.5 and 2 times solar, the mass
required to burn 50\% of the initial deuterium is between
$\sim$12.2~$M_J$ and $\sim$13.7~$M_J$.  Values somewhat outside this
range are not impossible, but values very far from this range are
probably rare.

\section{Conclusions}
\label{sec:conc}
We have calculated a suite of models of substellar mass objects,
encompassing a range of values of helium fraction, initial deuterium
fraction, and metallicity.  We have also used several calculations of
the atmospheric boundary conditions.  Although the rate of
deuterium-burning is extremely sensitive to temperature, it is
worthwhile bearing in mind that deuterium will burn at {\it some} rate
at any nonzero temperature, so one must specify what is meant by
``deuterium burning'' (i.e., how much of the initial deuterium must
burn to qualify).  We have found that, depending on what criterion is
used, reasonable values for the deuterium mass limit range from
$\sim$11.4 to $\sim$14.4 times Jupiter's mass; the former limit
corresponding to $2\times$ solar metallicity, 10\% of initial
deuterium burned, and the latter limit to solar metallicity, $Y=0.25$,
90\% of initial deuterium burned.  Extreme models (very low helium
fraction, very low metallicity) could require even higher mass in
order to burn a specified percentage of their initial deuterium.  The
canonical value of 13~$M_J$ \citep{burrows_et_al1997} is a reasonable
estimate of the mass at which signifcant deuterium burning begins, but
this value is model-dependent.

Still, one should keep in mind that the deuterium cut is probably less
relevant to an object's true taxonomic status than is its formation
history.  The downside of a ``formation scenario'' definition is that
formation history is not easily observable.  On the other hand, though
mass might be observable, the parameters $Y$, $y_{\rm D,i}$, and $Z$
might not be.  Furthermore, some objects will be found significantly
above 13~$M_J$ that will be of clearly ``planetary'' origin (e.g.,
CoRoT-3b).  Whether to call these objects ``brown dwarfs'' or
``deuterium-burning planets'' remains open for debate.  Similarly,
objects might be found that are significantly less massive than
13~$M_J$ that appear not to have formed in a protoplanetary disk, but
rather to be the low-mass end of the brown dwarf formation process.
Though classified as ``planets'' by standard terminology, these
objects might be more taxonomically related to brown dwarfs than to
planets.

We emphasize that there is really no need at this time to rigidly
distinguish between a giant planet and a brown dwarf on the basis of a
single criterion.  There is ambiguity in the provenance of these
objects, and this ambiguity might persist for a while.  The use of a
particular term (``planet'' or ``brown dwarf''), should be accompanied
by the definition that is employed.  Given that planets are thought to
be objects in orbit around a star (or around a brown dwarf), while
brown dwarfs are thought to be the low-mass end of the star formation
process, there is likely to be overlap in the mass range of these
objects unless one adopts a rigid mass cut to distinguish them.  Doing
so, however, will surely lead to an overlap in formation scenarios (as
in the case of CoRoT-3b).  When classifying a newly-discovered
substellar-mass object, one can use a variety of ``reasonable''
criteria, but the classifications should remain tentative until a more
thorough observational and theoretical understanding of substellar
objects is achieved.

\acknowledgments

We thank Kevin Heng, Bill Hubbard, Didier Saumon, Jason Nordhaus, and
Zimri Yaseen for useful discussions.  This study was supported in part
by NASA grant NNX07AG80G.  We also acknowledge support through
JPL/Spitzer Agreements 1328092, 1348668, and 1312647.

\bibliography{biblio.bib}

\clearpage

\begin{table}[t!]
\small
\begin{center}
\caption{Models}
\label{ta:models}
\begin{tabular}{l|cclcccl}
\hline
\hline
                        &           &                    &      &                  &                  & \\[0.2cm]
Model                   & $Z$       & {$y_{\rm D,i}$}     &  $Y$ & $M$ (10\%) & $M$ (50\%) & $M$ (90\%) \\[0.2cm]
                        & ($Z_\sun$\tablenotemark{a}) &  &      & ($M_J$)       & ($M_J$)       & ($M_J$) \\[0.2cm]
\hline
\rule {-3pt} {10pt}
He22                  & 1       &  $2\times 10^{-5}$ & 0.22       & 13.20 & 14.08 & 14.92 \\[0.2cm]
He25 (D2)\tablenotemark{b} & 1       &  $2\times 10^{-5}$ & 0.25  & 12.70 & 13.55 & 14.35 \\[0.2cm]
He28                  & 1       &  $2\times 10^{-5}$ & 0.28       & 12.06 & 12.83 & 13.62 \\[0.2cm]
He30                  & 1       &  $2\times 10^{-5}$ & 0.30       & 11.62 & 12.42 & 13.20 \\[0.2cm]
He32                  & 1       &  $2\times 10^{-5}$ & 0.32       & 11.32 & 12.05 & 12.78 \\[0.2cm]
  \tableline
                     &         &                    &       &     &      & \\[0.2cm]
D1                   & 1       &  $1\times 10^{-5}$ & 0.25        & 12.71 & 13.74 & 14.55 \\[0.2cm]
D2 (He25)            & 1       &  $2\times 10^{-5}$ & 0.25        & 12.70 & 13.55 & 14.35 \\[0.2cm]
D4                   & 1       &  $4\times 10^{-5}$ & 0.25        & 12.45 & 13.09 & 13.93 \\[0.2cm]
  \tableline
                     &         &                    &       &    &      & \\[0.2cm]
Z0                   & 0       &  $2\times 10^{-5}$ & 0.22        & 14.37 & 15.40 & 16.30 \\[0.2cm]
  \tableline
                     &         &                    &       &    &      & \\[0.2cm]
Z0.01                & 0.01    &  $2\times 10^{-5}$ & 0.22        & 13.59 & 14.56 & 15.39 \\[0.2cm]
Z0.1                 & 0.1     &  $2\times 10^{-5}$ & 0.22        & 12.92 & 13.82 & 14.65 \\[0.2cm]
Z0.3                 & 0.3     &  $2\times 10^{-5}$ & 0.22        & 12.48 & 13.33 & 14.14 \\[0.2cm]
Z1.0\tablenotemark{c} & 1       &  $2\times 10^{-5}$ & 0.22       & 12.00 & 12.79 & 13.56 \\[0.2cm]
  \tableline
                     &         &                    &      &     &      &  \\[0.2cm]
T0.3                 & $10^{-1/2}$ &  $1\times 10^{-5}$ & 0.25     & 12.92 & 13.77 & 14.67  \\[0.2cm]
T1                   & 1       &  $2\times 10^{-5}$ & 0.25        & 12.54 & 13.48 & 14.33  \\[0.2cm]
T3                   & $10^{1/2}$ &  $4\times 10^{-5}$ & 0.25      & 12.20 & 13.13 & 13.86  \\[0.2cm]
\end{tabular}
%
\vspace{0.1in}
\end{center}
\end{table}

{\small
\noindent {Models He22--He32 and D1--D4 are calculated with a
  \citet{burrows_et_al1997} radiative boundary condition Models
  Z0.01--Z1.0 are calculated with opacities from
  \citet{allard+hauschildt1995}, incorporated in atmospheric boundary
  conditions as described in \citet{burrows_et_al2001} (D. Saumon,
  private communication).  Models T0.3--T3 have radiative boundary
  conditions calculated with {\tt COOLTLUSTY}.  Model Z0's radiative
  boundary condition comes from \citet{saumon_et_al1994}.\\}
\\
 $^{\rm a}${Our adopted value of $Z_\sun = 0.0189$ is taken from
  \citet{anders+grevesse1989}.  This is in contrast to the 0.0142
  value from the more recent work of \citet{asplund_et_al2009}.  In
  absolute units, the quantities in the table correspond to the
  following values: $0\times Z_\sun = 0$; $0.01\times Z_\sun = 1.89
  \times 10^{-4}$; $0.3 \times Z_\sun = 0.00598$.\\}
\\
 $^{\rm b}${This model shows up in two rows, named both He25 and D2,
  and is one of two ``fiducial'' models.\\}
\\
 $^{\rm c}${This is the other ``fiducial'' model.\\}
}

\clearpage

\begin{table}[t]
\small
\begin{center}
\caption{Approximate Edge-Mass Derivatives}
\label{ta:derivs}
\begin{tabular}{l|cl}
\hline
\hline
                       &                             &                          \\[0.2cm]
 Model Sequence        & Derivative                  & Derivative Value  \\[0.2cm]
\hline
                       &                             &                          \\[0.2cm]
He22--He32             & $(dM/dY)/100$    &  $-0.2 M_J$             \\[0.2cm]
D1--D4                 & $(dM/d{y_{\rm D,i}})/10^5$  &  $-0.16 M_J$ \\[0.2cm]
D1--D4                 & $dM/d\log_{10}[{y_{\rm D,i}}]$ &  $-0.8 M_J$ \\[0.2cm]
Z0.01--Z1.0            & $dM/d\log_{10}[Z]$   &  $-0.9 M_J$            \\[0.2cm]
T0.3--T3               & $dM/d\log_{10}[Z]$   &  $-0.7 M_J$            \\[0.2cm]
  \tableline
\end{tabular}
\end{center}
\end{table}

\vspace{0.5in}

\begin{table}[h]
\small
\begin{center}
\caption{Edge-Masses for ``Realistic'' Treatment of $Z$}
\label{ta:redge} 
\begin{tabular}{c|ccc}
\hline
\hline
                       &          &          &          \\[0.2cm]
Metallicity            & $M$ (10\%) & $M$ (50\%) & $M$ (90\%)     \\[0.2cm]
                       &  ($M_J$)   & ($M_J$)    & ($M_J$)         \\[0.2cm]
\hline
            &          &          &          \\[0.2cm]
0.5$\times$ solar\tablenotemark{a} &  12.26   & 13.10    & 13.93    \\[0.2cm]
1 $\times$ solar       &  11.89   & 12.73    & 13.57    \\[0.2cm]
2 $\times$ solar       &  11.39   & 12.23    & 13.06    \\[0.2cm]
3 $\times$ solar       &  10.99   & 11.83    & 12.67    \\[0.2cm]
  \tableline
\end{tabular}
\end{center}
\end{table}

{\small
\noindent {We start with $Y=0.2703$ and add 0.0142 in $Y$ for every
  unit in $Z$.  We perform a cubic interpolation in $Y$ and use the
  derivatives recorded in the T0.3--T3 row of Table~\ref{ta:derivs}
  to approximate the combined affects of metallicity on pressure
  support and on atmospheric opacity.\\}
\\
 $^{\rm a}${Our adopted value of $Z_\sun = 0.0189$ is taken from
  \citet{anders+grevesse1989}.  In absolute units, the quantities in
  the table correspond to the following values: $0.5 \times Z_\sun =
  0.00945$; $2\times Z_\sun = 0.0378$; $3 \times Z_\sun = 0.0567$.\\}
}

\clearpage

\clearpage

\clearpage

\end{document}